\documentclass[10pt,amssymb,amsmath,aps,showpacs,floatfix,prl,nofootinbib]{revtex4-1}

\usepackage{graphicx}
\usepackage{xcolor}
\usepackage{appendix}

\begin{document}

\title{Quadrupole--octopole alignment of CMB related to primordial power spectrum with dipolar modulation in anisotropic spacetime}

\author{Zhe Chang$^{1,2}$\footnote{E-mail: changz@ihep.ac.cn}}
\author{Xin Li$^{1,2}$\footnote{E-mail: lixin@ihep.ac.cn}}
\author{Sai Wang$^{1}$\footnote{E-mail: wangsai@ihep.ac.cn}\footnote{Corresponding author.}}
\affiliation{${}^1$Institute of High Energy Physics\\ Chinese Academy of Sciences, 100049 Beijing, China\\
${}^2$Theoretical Physics Center for Science Facilities\\ Chinese Academy of Sciences, 100049 Beijing, China}


\begin{abstract}
The WMAP and Planck observations show that the quadrupole and octopole orientations of the CMB might align with each other. We reveal that the quadrupole--octopole alignment is a natural implication of the primordial power spectrum in an anisotropic spacetime. The primordial power spectrum is presented with a dipolar modulation. We obtain the privileged plane by employing the ``power tensor'' technique. At this plane, there is the maximum correlation between quadrupole and octopole. The probability for the alignment is much larger than what in the isotropic universe. We find that this model would lead to deviations from the statistical isotropy only for low--\(\ell\) multipoles.
\end{abstract}
\maketitle

\section{I. Introduction}
\label{introduction}
The universe is assumed to be statistically isotropic at large scales in the standard model of cosmology.
Recently, however, several evidences appeared for possible deviations from
the statistical isotropy of the CMB temperature fluctuations \cite{WMAP9001,Planck2013resultsXXIII}.
One refers the alignment of the quadrupole and octopole orientations.
Based on the one-year WMAP data \cite{WMAP1001}, Tegmark et al. \cite{quadrupole-octopole alignment}
and Copi et al. \cite{quadrupole-octopole alignment by Schwarz,quadrupole-octopole alignment by Copi1,quadrupole-octopole alignment by Copi}
originally proposed that there is a quadrupole--octopole alignment of CMB anisotropy.
The recent nine-year WMAP results \cite{WMAP9001} also show that the quadrupole--octopole alignment is of about \(3^{\circ}\).
The Planck 2013 results \cite{Planck2013resultsXXIII} confirm the evidence for this alignment
while there is a misalignment between the quadrupole and octopole orientations by an amount between \(9^{\circ}\) and \(13^{\circ}\).
However, the quadrupole and octopole orientations are distributed uniformly in the isotropic universe.
The possibility for the quadrupole--octopole alignment is expected to be much smaller.

The significance of the quadrupole--octopole alignment based on the Planck dataset becomes smaller than what based on the WMAP dataset.
Nevertheless, we cannot simply drop the proposition that the universe might be statistically anisotropic at large scales.
The reason is that there are several other ``anomalies'' of the CMB anisotropy at low--\(\ell\) multipoles.
For instance, the hemispherical asymmetry \cite{hemispherical asymmetry01,hemispherical asymmetry02,hemispherical asymmetry03},
the parity asymmetry \cite{parity asymmetry01,parity asymmetry02,parity asymmetry03,parity asymmetry04,parity asymmetry05,parity asymmetry06,Dipole01,Dipole02,Mirror parity01,Mirror parity02,Mirror parity03,Directional dependence of CMB parity asymmetry},
the dipolar power modulation \cite{Spontaneous isotropy breaking,dipolar power modulation01,dipolar power modulation02,dipolar power modulation03},
and so on.
These anomalies might be related to the deviation from isotropy if they occupy certain cosmological origins \cite{Planck2013resultsXXIII}.
There is a useful approach called the ``power tensor'' technique \cite{The Virgo Alignment Puzzle in Propagation of Radiation on Cosmological Scales,Testing Isotropy of Cosmic Microwave Background Radiation,Direction dependence of the power spectrum1302}
to study the deviations from the isotropy in the CMB.
The power tensor is a generalization of the power spectrum.
In the limit of isotropy, its ensemble expectation would reduce to the power spectrum.
However, it could not be diagonal in the anisotropic universe.
The correlation between two power tensors directly describes the level of alignment between two multipoles.
In this paper, we employ this method to study the quadrupole--octopole alignment in an anisotropic spacetime.

The deviations from isotropy may be related with a privileged axis in the universe.
We have proposed an anisotropic inflation model in the Randers spacetime
to study the dipole modulation of the CMB temperature fluctuations \cite{Inflation and primordial power spectra at anisotropic spacetime}.
In the Randers spacetime \cite{Randers space}, the line element includes an extra one-form in addition to the Riemannian line element.
The one-form would reduce the number of Killing vectors.
Locally, this one-form has been extensively studied in very special relativity (VSR) \cite{VSR}.
The local Randers metric was proved to possess symmetries of the group \(TE(2)\) \cite{Finsler isometry LiCM,The Finsler Type of Space-time Realization of Deformed Very Special Relativity}.
Thus, the Randers spacetime has less symmetries than the Riemannian Friedmann-Robertson-Walker (FRW) spacetime does.
There is a privileged axis in the Randers spacetime.
It would induce the primordial power spectrum with a dipolar modulation if the inflationary universe is of Randers type \cite{Inflation and primordial power spectra at anisotropic spacetime}.
This initial spectrum with anisotropy is compatible with the Planck 2013 results.

The Randers spacetime belongs to Finsler geometry  \cite{Finsler,Book by Bao}.
Finsler geometry gets rid of the quadratic restriction on the line element.
It is a natural generalization of Riemann geometry and includes Riemann geometry as a special case.
It is intrinsically anisotropic.
The isometric transformation shows that there are less symmetries in Finsler geometry than in Riemann geometry \cite{Finsler isometry LiCM,Finsler isometry by Wang,Finsler isometry by Rutz}.
Thus, Finsler geometry could be a reasonable candidate to account for the deviations from isotropy at large scales in the universe.
In this paper, we employ the Randers--Finsler spacetime to study the quadrupole--octopole alignment.
The anisotropic inflation model and the primordial power spectrum with dipolar modulation are introduced briefly.
By the ``power tensor'' technique, we study the correlations between two multipoles.
We obtain eigenvectors with the maximum eigenvalue of the correlation matrix of two power tensors.
Several quantities, which describes the correlations between two multipoles, are evaluated.

The rest of the paper is arranged as follows.
In Section II, we introduce the ``power tensor'' technique to study the statistical anisotropy.
The correlation matrix of two power tensors is defined to study the statistical correlation between two multipoles.
In Section III, we briefly review the inflation model with anisotropy in the Randers--Finsler spacetime.
The primordial power spectrum is presented with a dipolar modulation.
In Section IV, we study the implications of the primordial power spectrum with anisotropy on the mode alignments of CMB anisotropy,
especially the quadrupole--octopole alignment.
Conclusions and remarks are listed in Section V.
In addition, there are two appendixes for calculating physical quantities explicitly.

\section{II. The ``power tensor'' technique}
\label{power tensor technique}
Generically, the CMB temperature fluctuation \(\frac{\delta T}{T}(\hat{\textbf{p}},\textbf{x},\eta)\)
could be decomposed in terms of the spherical harmonics \(Y_{\ell m}(\hat{\textbf{p}})\), namely,
\begin{equation}
\label{CMB perturbation into spherical harmonics 1}
\frac{\delta T}{T}(\hat{\textbf{p}},\textbf{x},\eta)=\sum_{\ell=1}^{\infty}\sum_{m=-\ell}^{\ell}a_{\ell m}Y_{\ell m}(\hat{\textbf{p}})\ ,
\end{equation}
where \(\hat{\textbf{p}}\) denotes the observed direction by an observer at the location \(\textbf{x}\), and \(\eta\) denotes the conformal time.
We assume that the CMB anisotropy satisfies the Gaussian distribution,
even though the statistical anisotropy might be related with the non-Gaussianity \cite{A Hemispherical Power Asymmetry from Inflation,The CMB asymmetry from inflation,Small non-Gaussianity and dipole asymmetry in the CMB,CMB dipole asymmetry from a fast roll phase,Large Scale Anisotropic Bias from Primordial non-Gaussianity,Hemispherical Asymmetry and Local non-Gaussianity a Consistency Condition,Asymmetric Sky from the Long Mode Modulations,Anisotropic Non-Gaussianity from a Two-Form Field,CMB Power Asymmetry from Primordial Sound Speed Parameter}.
The variance of harmonic coefficients \(a_{\ell m}\) gives the two-point correlator.
It is given by
\begin{equation}
\label{power spectrum--two point correlator}
\langle a_{\ell m}a^{*}_{\ell' m'}\rangle=C_{\ell\ell' m m'}\ ,
\end{equation}
where the angle brackets stand for the ensemble average.
The correlators \(C_{\ell\ell' m m'}\) might be not diagonal in the anisotropic case.
In the case of statistical isotropy, it reduces to the power spectrum \(C_{\ell\ell' m m'}=C_{\ell}\delta_{\ell \ell'}\delta_{m m'}\).

At the largest scales, the CMB temperature fluctuations are linear.
They could be given by
\begin{equation}
\label{CMB perturbation given by primordial condition}
\frac{\delta T}{T}(\hat{\textbf{p}},\textbf{x},\eta)=\int \frac{d^3\textbf{k}}{(2\pi)^{3/2}}\mathcal{R}(\textbf{k},\eta_{i})\Delta(k,\hat{\textbf{k}}\cdot\hat{\textbf{p}},\eta)e^{i\textbf{k}\cdot\textbf{x}}\ ,
\end{equation}
where \(k=|\textbf{k}|\), and \(\mathcal{R}(\textbf{k},\eta_{i})\) is the initial curvature perturbation at the initial time \(\eta_{i}\).
Here \(\Delta(k,\hat{\textbf{k}}\cdot\hat{\textbf{p}},\eta)\) denotes the transfer function,
which characterizes the change of amplitude of the CMB temperature fluctuation from the conformal time \(\eta_{i}\) to today \(\eta\).
It could be expanded in terms of Legendre polynomials,
\begin{equation}
\label{transfer function}
\Delta(k,\hat{\textbf{k}}\cdot\hat{\textbf{p}},\eta)=\sum_{\ell}(-i)^{\ell}\left(\frac{2\ell+1}{4\pi}\right)
P_{\ell}(\hat{\textbf{k}}\cdot\hat{\textbf{p}})\Delta_{\ell}(k,\eta)\ .
\end{equation}
According to Eq.~(\ref{CMB perturbation into spherical harmonics 1})--(\ref{transfer function}),
the correlators \(C_{\ell\ell' m m'}\) could be calculated as \footnote{See \ref{A} for an explicit calculation.}
\begin{equation}
\label{two point correlator}
C_{\ell\ell' m m'}=(-i)^{\ell-\ell'}\int{d^3\textbf{k}}\frac{2\pi^{2}}{k^{3}}\mathcal{P}_{\mathcal{R}}(\textbf{k})
\Delta_{\ell}(k,\eta)\Delta^{*}_{\ell'}(k,\eta)Y^{*}_{\ell m}(\hat{\textbf{k}})Y_{\ell' m'}(\hat{\textbf{k}})\ .
\end{equation}
Here, the primordial power spectrum \(\mathcal{P}_{\mathcal{R}}(\textbf{k})\) is given by
\begin{equation}
\label{primordial power spectrum}
\langle\mathcal{R}(\textbf{k})\mathcal{R}^{*}(\textbf{k}')\rangle=
\delta^{(3)}(\textbf{k}-\textbf{k}')\frac{2\pi^2}{k^{3}}\mathcal{P}_{\mathcal{R}}(\textbf{k})\ ,
\end{equation}
which encodes all the properties of statistical anisotropy.

In this paper, we employ the ``power tensor'' technique \cite{The Virgo Alignment Puzzle in Propagation of Radiation on Cosmological Scales,Testing Isotropy of Cosmic Microwave Background Radiation,Direction dependence of the power spectrum1302}
to study the possible statistical anisotropy of the CMB temperature fluctuations in an anisotropic spacetime.
The power tensor is defined by a second-rank matrix as
\begin{equation}
\label{power tensor}
A_{ij}(\ell)=N(\ell)\sum_{m,m',m''}\langle \ell m|J_{i}|\ell m'\rangle\langle \ell m''|J_{j}|\ell m\rangle a_{\ell m'}a^{*}_{\ell m''}\ ,
\end{equation}
where the coefficient \(N(\ell)\) refers \footnote{This is slightly different from the original definition in
Ref.~\cite{Testing Isotropy of Cosmic Microwave Background Radiation,Direction dependence of the power spectrum1302}.
One should note that the power tensor requires the condition \(\langle A_{ij}(\ell)\rangle=\frac{1}{3}C_{\ell}\delta_{ij}\)
in the limit of statistical isotropy. The definition here satisfies this condition.}
\begin{equation}
\label{Nl}
N(\ell)=\frac{1}{\ell(\ell+1)(2\ell+1)}\ .
\end{equation}
Here, \(J_{i}\) denotes the \(i\)-th component of angular momentum operator,
and the state \(|\ell m\rangle\) stands for the spin-\(\ell\) representation.
The power tensor is real and symmetric.
The privileged direction of the \(\ell\)-multipole is given by the eigenvector of \(\langle A_{ij}(\ell)\rangle\) with the maximum eigenvalue.
However, it is possible that \(\langle A_{ij}(\ell)\rangle\) has no preferred directions
when the statistical anisotropy has certain particular behaviors.
For instance, the initial power spectrum \(\mathcal{P}_{\mathcal{R}}(\textbf{k})\) may acquire a dipolar modulation
\cite{Direction dependence of the power spectrum1302,Inflation and primordial power spectra at anisotropic spacetime}.
The two-point correlators could reduce to the isotropic form, since a dipole is point-parity asymmetric, see \ref{A}.
In this case, the multipoles would acquire preferred directions occasionally.
One should note that the privileged direction of a multipole is not statistically significant.
Only the alignment of two multipoles reveals the statistical anisotropy possibly.

There is a useful way to describe the alignment of two multipoles.
It refers to the correlator of two power tensors with different multipoles \cite{Direction dependence of the power spectrum1302}.
Thus, the correlation of two power tensors is defined as
\footnote{This form of the correlation has slight difference from the original one in Ref.~\cite{Direction dependence of the power spectrum1302}.
Here, we have only extracted the anisotropic effects.
In the original definition, however, \(S(\ell,\ell')\) is given by the correlator of two power tensors.
Thus, the isotropic effects are still contained in the original definition.
One should note that these two definitions would point to the same physical results.}
\begin{equation}
\label{power tensor correlation trace}
S(\ell,\ell')=\frac{3}{C_{\ell}C_{\ell'}}Tr\left(\langle A^{+}(\ell)A(\ell')\rangle_{ij}\right)-1\ .
\end{equation}
This quantity describes the level of alignment between two multipoles.
It is directly a statistically significant quantity.
In the isotropic case, \(S(\ell,\ell')\) would vanish.
Even though \(S(\ell,\ell')\) could give the magnitude of correlations, it is direction-blind
since it cannot determine the direction (or the plane) of the maximum correlation.
Statistically, however, the alignment of two multipoles would lie on the direction(s) of the maximum correlation.
Fortunately, it is easy to overcome the above problem.
One could define a second-rank real symmetric matrix as
\begin{equation}
\label{power tensor correlation matrix}
S_{ij}(\ell,\ell')=\frac{3}{C_{\ell}C_{\ell'}}\langle A^{+}(\ell)A(\ell')\rangle_{ij}-\frac{1}{3}\delta_{ij}\ .
\end{equation}
In the following, we call this matrix the ``correlation matrix'' of power tensors, or ``correlation matrix'' for short.
We could obtain the eigenvalues and the related eigenvectors of \(S_{ij}(\ell,\ell')\).
The eigenvectors with the maximum eigenvalue would give the privileged directions, which refer to the maximum correlation of the two multipoles.
In Section IV, we will follow this approach to study the quadrupole--octopole alignment of the CMB anisotropy in an anisotropic spacetime.

\section{III. Primordial power spectra with anisotropy}
\label{primordial power spectrum with anisotropy}
Recently, an anisotropic inflation model has been proposed explicitly
in Ref.~\cite{Inflation and primordial power spectra at anisotropic spacetime}.
This model suggested that the spacetime of very early universe may be of Randers--Finsler type.
The Friedmann equation as well as its inflationary solution was presented at first order.
The primordial power spectrum was obtained with direction dependence.
For example, it might acquire a dipolar modulation at the first--order approximation.
In this section, we briefly review the main results of the anisotropic inflation model.

Finsler geometry \cite{Finsler,Book by Bao} is defined by the Finsler structure \(F(x,y)\),
where \(x\) denotes a position and \(y=dx/d\tau\) a fibre coordinate.
The Finsler structure is a smooth positive function on the tangent bundle.
It satisfies the property of positively homogeneity of degree one,
\begin{equation}
\label{Finsler structure}
F(x,\lambda y)=\lambda F(x,y),~~\rm{for}~\lambda>0\ .
\end{equation}
The Finsler metric is defined by
\begin{equation}
\label{Finsler metric}
g_{\mu\nu}=\frac{\partial}{\partial y^{\mu}}\frac{\partial}{\partial y^{\nu}}\left(\frac{1}{2}F^{2}\right)\ .
\end{equation}
The Randers spacetime \cite{Randers space} is a class of Finsler spacetime.
The Randers(--Finsler) structure is given as
\begin{equation}
\label{Randers structure}
F(x,y)=\alpha(x,y)+\beta(x,y)\ ,
\end{equation}
where
\begin{eqnarray}
\alpha(x,y)&=&\sqrt{\tilde{a}_{\mu\nu}(x)y^{\mu}y^{\nu}}\ ,\\
\beta(x,y)&=&\tilde{b}_{\mu}(x)y^{\mu}\ .
\end{eqnarray}
Here \(\alpha\) denotes a Riemann structure and \(\tilde{a}_{\mu\nu}\) the Riemann metric.
\(\beta\) denotes a 1-form, which may be related to a vector field.
In the following discussions, we choose \(\tilde{a}_{\mu\nu}\) the spatially flat Friedmann-Robertson-Walker (FRW) metric,
i.e., \(\tilde{a}_{\mu\nu}=\rm{diag}(1,-a^2(t),-a^2(t),-a^2(t))\).
The 1-form \(\beta\) has just the temporal component
\begin{equation}
\tilde{b}_{\mu}=B(x^3)\delta^{0}_{\mu}\ ,
\end{equation}
where \(B(x^3)\) is a function of only the third spatial coordinate \(x^3\).

In the osculating Riemannian approach \cite{Book by Rund,Book by Asanov,FRW model with weak anisotropy by Stavrinos},
the Finsler structure could be related with certain osculating Riemannian metric,
namely, \(g_{\mu\nu}(x)=g_{\mu\nu}(x,y(x))\).
The fibre coordinate \(y\) could be viewed as a function of the spacetime position \(x\).
We are only interested in the evolution of the very early universe.
After our calculation of the connection and the curvature tensor,
Einstein's gravitational equation could be obtained as
\cite{Inflation and primordial power spectra at anisotropic spacetime,FRW model with weak anisotropy by Stavrinos}
\begin{equation}
\label{Einstein's field equations}
Ric_{\mu\nu}-\frac{1}{2}g_{\mu\nu}S=8\pi G T_{\mu\nu}\ ,
\end{equation}
where \(Ric_{\mu\nu}\) denotes the Ricci tensor, \(S\) denotes the scalar curvature, and
\begin{equation}
T_{\mu\nu}(\phi)=\partial_\mu \phi\,\partial_\nu \phi \,-\,
g_{\mu\nu}\left( \frac{1}{2}\,g^{\alpha\beta}\,\partial_\alpha
\phi\,\partial_\beta \phi \,-\, V(\phi)\right)
\end{equation}
is the energy-momentum tensor of the inflaton field. Here \(V(\phi)\) is a potential of the inflaton.

In the very early time, the universe might undergo an era of exponential expansion,
which is called inflation \cite{Inflation by Starobinsky,Inflation by Guth,Inflation by Linde,Inflation by Steinhardt,Inflation by Linde0}.
The scale factor expands as \(a(t)\sim e^{Ht}\) where \(H\) denotes the Hubble constant.
The inflaton could be decomposed as \(\phi=\phi^{(0)}(t)+\delta \phi(t,\textbf{x})\).
Here \(\phi^{(0)}(t)\) denotes the zero--order part which drives the inflation.
The \(00\) component of Einstein's gravitational equation gives the Friedmann equation
\cite{Inflation and primordial power spectra at anisotropic spacetime}
\begin{equation}
\label{Friedmann equation}
3\left(\frac{\dot{a}}{a}\right)^2+\left(\frac{1}{a}\right)^2\frac{3B'^2-4B''(1+B)}{4(1+B)}
=8\pi GT_{00}(\phi^{(0)})\ ,
\end{equation}
where the dot and the prime denote derivatives with respective to \(t\) and \(x^3\), respectively.
At first order, the above Friedmann equation could have an inflationary solution
\cite{Inflation and primordial power spectra at anisotropic spacetime}
\begin{eqnarray}
a(t)&\propto& e^{Ht}\ ,\\
\label{B(x3)}
B(x^3)&=&r_{c}^{-1}x^{3}+\mathcal{O}(x^{3})\ ,
\end{eqnarray}
where \(r_{c}\) stands for a spatial scale for cutoff.
This cutoff scale would be comparable to the observable scale of the universe.
One should note that \(B(x^{3})\) depends only \(x^3\) such that the space is cylindrically symmetric instead of spherically symmetric.
Thus, the universe undergone a phase of anisotropic inflation.

The anisotropic inflation should leave certain anisotropic imprints on the observable universe \cite{Cosmological Magnetic Fields05}.
The primordial power spectrum has been calculated in Ref.~\cite{Inflation and primordial power spectra at anisotropic spacetime}.
It could acquire a dipolar modulation besides the isotropic power spectrum.
To obtain the primordial power spectrum, one should study the equation of motion for
the first--order perturbation \(\delta\phi(t,\textbf{x})\) of the inflaton
\cite{Fluctuation01,Fluctuation02,Fluctuation03,Fluctuation04,Fluctuation05,Fluctuation06,Fluctuation07}, see also Ref.~\cite{Particle physics models of inflation and curvaton scenarios,Visible sector inflation and the right thermal history in light of Planck data}.
The equation of motion for \(\delta\phi(t,\textbf{x})\) could be given as \cite{Inflation and primordial power spectra at anisotropic spacetime}
\begin{equation}
\label{equation of motion first order}
\ddot{\delta\phi}+3H\dot{\delta\phi}-\frac{1+B}{a^2}\nabla^2\delta\phi-\frac{3}{2}\frac{B'}{a^2}\delta\phi'=0\ .
\end{equation}
In the momentum space, this equation becomes
\begin{equation}
\label{Fluctuations motion Fourier modes}
\ddot{\delta\phi_{\textbf{k}}}+3H\dot{\delta\phi_{\textbf{k}}}+\frac{k_{eff}^{2}}{a^2}\delta\phi_{\textbf{k}}=0\ ,
\end{equation}
where the effective wavenumber \(k_{eff}\) is given by
\begin{equation}
\label{keff}
k_{eff}^{2}=k^{2}\left(1+B-i\frac{3B'}{2k}(\hat{\textbf{k}}\cdot\hat{\textbf{x}}^3)\right)\ .
\end{equation}
At the super-horizon scales, therefore, we obtain the primordial power spectrum of the form
\cite{Inflation and primordial power spectra at anisotropic spacetime}
\begin{equation}
\label{primordial power spectra of phi approximate}
\mathcal{P}_{\delta\phi_{\textbf{k}}}=\left(\frac{H}{2\pi}\right)^{2}\left(\frac{k}{|k_{eff}|}\right)^{3}\ .
\end{equation}
As a first--order approximation, thus, the initial power spectrum (\ref{primordial power spectrum})
of comoving curvature perturbation could be obtained as
\begin{equation}
\label{Primordial power spectra of R}
\mathcal{P}_{\mathcal{R}}\simeq\left(\frac{H}{\dot{\phi}}\right)^{2}\mathcal{P}_{\delta\phi_{\textbf{k}}}
\simeq\mathcal{P}_{\mathcal{R}}^{iso}(k)\left(1-3B\right)\ ,
\end{equation}
where \(\mathcal{P}_{\mathcal{R}}^{iso}(k)\simeq \left(\frac{H}{2\pi}\right)^{2}\).
We have dropped the terms of higher orders in \(B\).
By using Eq.~(\ref{B(x3)}), the initial power spectrum (\ref{Primordial power spectra of R}) could be rewritten as
\begin{equation}
\label{primordial power spectra of R final form}
\mathcal{P}_{\mathcal{R}}({\textbf{k}})=\mathcal{P}_{\mathcal{R}}^{iso}(k)\left(1+\frac{k_{c}}{k}(\hat{\textbf{k}}\cdot\hat{\textbf{n}})\right)\ .
\end{equation}
We have used the relation \(x\sim k^{-1}\) in the above equation, and extra constants have been absorbed into \(k_{c}\).
Here \(\hat{\textbf{n}}\equiv \hat{\textbf{x}}^3\) denotes the privileged axis in general.
The parameter \(k_{c}\) denotes a critical wavenumber, which refers to the range of the statistical anisotropy.
By comparing (\ref{primordial power spectra of R final form}) with (\ref{g(k)}),
we obtain a relation \(g(k)=k_{c}k^{-1}\).
This prediction implies that the anisotropic effects would be significant at large scales only.
It is compatible with the observable constraints at present \cite{Inflation and primordial power spectra at anisotropic spacetime,The CMB asymmetry from inflation}.

\section{IV. Implications on the quadrupole--octopole alignment}
\label{mode alignment}
In the above section, we have briefly reviewed the inflation model and the anisotropic power spectrum in Randers spacetime.
It was revealed that the power spectrum acquires a dipolar modulation.
In this section, we apply this dipole--modulated power spectrum to study the quadrupole--octopole alignment of the CMB anisotropy.

We are only interested in the case \(\ell'=\ell+1\).
The initial power spectrum has been presented in Eq.~(\ref{primordial power spectra of R final form}).
Thus, the correlators \(C_{\ell \ell' m m'}\)'s in Eq.~(\ref{two point correlator}) could be computed as
\begin{eqnarray}
\label{all Cl}
C_{\ell \ell' m m'}&=&C_{\ell \ell' m m'}^{iso}+C_{\ell \ell' m m'}^{aniso}\ ,\\
\label{iso}
C_{\ell \ell' m m'}^{iso}&=&\delta_{\ell\ell'}\delta_{mm'}\int \frac{dk}{k}\mathcal{P}_{\mathcal{R}}^{iso}(k)\Delta_{\ell}(k,\eta)\Delta_{\ell}^{*}(k,\eta)\ ,\\
\label{aniso}
C_{\ell \ell' m m'}^{aniso}&=&(-i)^{\ell-\ell'+1}\zeta_{\ell m; \ell' m'}\int \frac{dk}{k}\mathcal{P}_{\mathcal{R}}^{iso}(k)\left(\frac{k_{c}}{k}\right)\Delta_{\ell}(k,\eta)\Delta_{\ell'}^{*}(k,\eta)\ .
\end{eqnarray}
Eq.~(\ref{iso}) corresponds to the isotropic case, while Eq.~(\ref{aniso}) describes  violations of the statistical isotropy.
Here, the coefficient \(\zeta_{\ell m; \ell' m'}\) is given by
\begin{eqnarray}
\label{zeta}
\zeta_{\ell m; \ell' m'}&=&\sqrt{\frac{4\pi}{3}} \int d\Omega_{\hat{\textbf{k}}}Y_{10}(\hat{\textbf{k}})Y^{*}_{\ell m}(\hat{\textbf{k}})Y_{\ell' m'}(\hat{\textbf{k}})\ \\
\label{zeta expansion}
&=&\delta_{m',m}\left(\delta_{\ell',\ell+1}\sqrt{\frac{(\ell-m+1)(\ell+m+1)}{(2\ell+1)(2\ell+3)}}
+\delta_{\ell',\ell-1}\sqrt{\frac{(\ell-m)(\ell+m)}{(2\ell+1)(2\ell-1)}}\right)\ .
\end{eqnarray}
There is a useful relation for the CMB anisotropy, namely, \(a_{\ell m}=(-1)^{m}a^{*}_{\ell -m}\).
Therefore, we could obtain \(\bar{C}_{\ell \ell' m m'}\equiv \langle a_{\ell m}a_{\ell' m'} \rangle\),
\begin{eqnarray}
\label{all Cl bar}
\bar{C}_{\ell \ell' m m'}&=&\bar{C}_{\ell \ell' m m'}^{iso}+\bar{C}_{\ell \ell' m m'}^{aniso}\ ,\\
\label{iso bar}
\bar{C}_{\ell \ell' m m'}^{iso}&=&(-1)^{m}\delta_{\ell\ell'}\delta_{mm'}\int \frac{dk}{k}\mathcal{P}_{\mathcal{R}}^{iso}(k)\Delta_{\ell}(k,\eta)\Delta_{\ell}^{*}(k,\eta)\ ,\\
\label{aniso bar}
\bar{C}_{\ell \ell' m m'}^{aniso}&=&(-1)^{m}(-i)^{\ell-\ell'+1}\bar{\zeta}_{\ell m; \ell' m'}\int \frac{dk}{k}\mathcal{P}_{\mathcal{R}}^{iso}(k)\left(\frac{k_{c}}{k}\right)\Delta_{\ell}(k,\eta)\Delta_{\ell'}^{*}(k,\eta)\ .
\end{eqnarray}
Here \(\bar{\zeta}_{\ell m;\ell' m'}\) could be obtained via replacing \(\delta_{m',m}\) in Eq.~(\ref{zeta expansion}) by \(\delta_{m',-m}\).

By a not too tedious calculation, we get the correlation matrix of power tensors in Eq.~(\ref{power tensor correlation matrix})
\begin{equation}
\label{correlation matrix 1}
S_{ij}(\ell,\ell')=\frac{3N(\ell)N(\ell')}{C_{\ell}C_{\ell'}}\sum_{m_{n}=-\ell}^{\ell}\sum_{m'_{n}=-\ell'}^{\ell'}
J_{ij} J
\left(\langle a^{*}_{\ell m_{2}} a_{\ell' m'_{2}} \rangle \langle a_{\ell m_{3}} a^{*}_{\ell'm'_{3}} \rangle +
\langle a^{*}_{\ell m_{2}} a^{*}_{\ell'm'_{3}} \rangle\langle a_{\ell m_{3}} a_{\ell' m'_{2}} \rangle\right)\ ,
\end{equation}
where
\begin{eqnarray}
\label{Jij}
J_{ij}&\equiv& \langle \ell m_{3}|J_{i}|\ell m_{1} \rangle^{*} \langle \ell' m'_{3}|J_{j}|\ell' m'_{1} \rangle\ ,\\
\label{J}
J&\equiv&\sum_{k}\langle \ell m_{1}|J_{k}|\ell m_{2} \rangle^{*} \langle \ell' m'_{1}|J_{k}|\ell' m'_{2} \rangle\ .
\end{eqnarray}
The explicit expressions of \(J\) and \(J_{ij}\) could be found in Appendix~B\ref{B}.
Here, we have used the expression of four-point correlators of \(a_{\ell m}\)'s as
\begin{equation}
\label{four point correlator}
\langle a^{*}_{\ell m_{2}} a_{\ell m_{3}} a_{\ell' m'_{2}} a^{*}_{\ell'm'_{3}} \rangle=
\langle a^{*}_{\ell m_{2}} a_{\ell m_{3}}\rangle \langle a_{\ell' m'_{2}} a^{*}_{\ell'm'_{3}} \rangle+
\langle a^{*}_{\ell m_{2}} a_{\ell' m'_{2}} \rangle \langle a_{\ell m_{3}} a^{*}_{\ell'm'_{3}} \rangle+
\langle a^{*}_{\ell m_{2}} a^{*}_{\ell'm'_{3}} \rangle\langle a_{\ell m_{3}} a_{\ell' m'_{2}} \rangle\ .
\end{equation}
Note that the term \(\langle a^{*}_{\ell m_{2}} a_{\ell m_{3}}\rangle \langle a_{\ell' m'_{2}} a^{*}_{\ell'm'_{3}} \rangle\)
would not induce the statistical anisotropy in the case of the dipole--modulated initial power spectrum, see \ref{A}.
It is easy to check that its effect cancels the isotropic term \(\frac{1}{3}\delta_{ij}\)
at the right hand side of Eq.~(\ref{power tensor correlation matrix}).
Thus, one need calculate only the other two terms at the right hand side of Eq.~(\ref{four point correlator}).
Finally, the correlation matrix (\ref{power tensor correlation matrix}) (or (\ref{correlation matrix 1})) could be given as
\begin{eqnarray}
\label{correlation matrix 2}
S_{ij}(\ell,\ell')=&&\frac{3N(\ell)N(\ell')}{C_{\ell}C_{\ell'}}\left(\sum_{m m'}J_{ij} J
\left(\zeta_{\ell m_{2};\ell' m'_{2}}\zeta^{*}_{\ell m_{3}; \ell' m'_{3}}+
\bar{\zeta}_{\ell m_{2};\ell' m'_{3}}\bar{\zeta}^{*}_{\ell m_{3}; \ell' m'_{2}}\right)\right)\nonumber\\
&&\times{\left(\int \frac{d k}{k}\mathcal{P}_{\mathcal{R}}^{iso}(k)\left(\frac{k_{c}}{k}\right)\Delta_{\ell}(k,\eta)\Delta_{\ell'}^{*}(k,\eta)\right)^2}\ .
\end{eqnarray}
At large scales, the transfer function is approximately given by \(\Delta_{\ell}(k)\propto J_{\ell}(k(\eta-\eta_{ls}))\approx J_{\ell}(k\eta)\).
Here, \(\eta\) and \(\eta_{ls}\) denote the conformal time today and at the last scattering, respectively.
The last approximation arises from \(\eta_{ls}\ll\eta\).
We could compute all the components of \(S_{ij}(\ell, \ell')\) for a given \(\ell\).
The eigenvulues and eigenvectors could also be calculated directly as was mentioned in Section II.
Thus, the privileged axis is singled out for the maximum correlation of the two multipoles.

We are interested in the quadrupole--octopole alignment of CMB anisotropy.
After a lengthy computation, it is showed that the correlation matrix \(S_{ij}(2,3)\) is diagonal in the coordinates chosen in this paper.
Analytically, Eq.~(\ref{correlation matrix 2}) for \(\ell=2\) is calculated as
\begin{equation}
\label{c2-c3 alignment}
S_{ij}(2,3)\simeq\left(k_{c}\eta\right)^2 \left(\begin{array}{ccc}
                                            0.0652 & 0 & 0 \\
                                            0 & 0.0652 & 0 \\
                                            0 & 0 & 0.0235
                                          \end{array}\right)\ .
\end{equation}
As was mentioned above, \(k_{c}\) denotes the critical wavenumber of the statistical anisotropy,
and \(\eta\) is the conformal time of the universe today.
Only when the critical scale \(k_{c}^{-1}\gg \eta\), the statistical isotropy recovers in the observable universe.
However, the observations of CMB anisotropy from WMAP and Planck
showed certain possible evidence for the statistical anisotropy at large scales.
This reveals that the critical scale should be comparable with the present scale of the observable universe,
namely, \(k_{c}\eta \sim 1\).

There is the other significant issue which refers the privileged direction of the quadrupole--octopole alignment.
From Eq.~(\ref{c2-c3 alignment}), the correlation matrix acquires the maximum eigenvalue along both \(x^{1}\) and \(x^{2}\) axes.
This result reveals that the quadrupole and the octopole have the maximum correlation
in the plane perpendicular to the privileged axis \(\hat{\textbf{n}}\).
Thus, the quadrupole and octopole orientations should lie in the plane with the uniform distribution.
We cannot determine which direction is preferred in the plane.
Actually, the nine-year WMAP data show that there is a \(3^{\circ}\) misalignment between the quadrupole and the octopole \cite{WMAP9001}.
The Planck 2013 results show a larger misalignment of the amount between \(9^{\circ}\) and \(13^{\circ}\) \cite{Planck2013resultsXXIII}.
Statistically, all the directions at this plane are equally likely in this model.
In the isotropic universe, however, the quadrupole and octopole orientations are distributed uniformly on the celestial sphere.
Thus, there is a much larger possibility for the quadrupole--octopole alignment in this anisotropic model than that in the isotropic model.

\begin{figure*}[h]
\begin{center}
\includegraphics[width=9 cm]{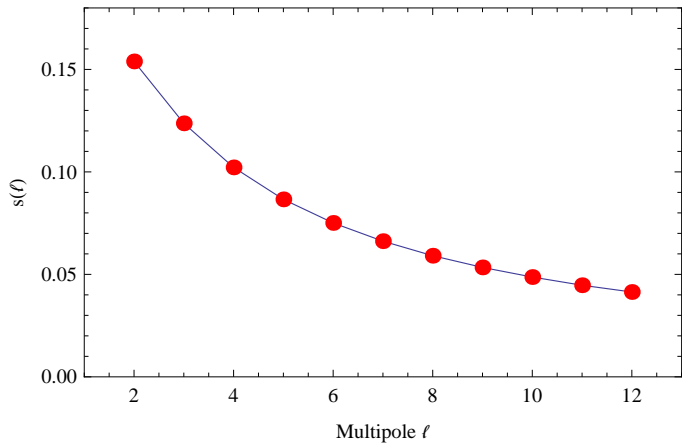}
\caption{\(s(\ell)\) is a decreasing function of multipoles \(\ell\)'s. [Color online]}
\label{sl}
\end{center}
\end{figure*}

Another significant issue refers to the decay rate of the anisotropic effect.
To describe this issue, we could compute the trace of correlation matrix \(S_{ij}(\ell,\ell+1)\).
The correlation \(S(\ell,\ell+1)\) could be written in the form
\begin{equation}
\label{Sll1}
S(\ell,\ell+1)=s(\ell)(k_{c}\eta)^2\ ,
\end{equation}
where \(s(\ell)\) describes the magnitude of the correlation between the two multipoles directly.
The numerical computation shows that \(s(\ell)\) is a decreasing function of \(\ell\), see Fig.~\ref{sl}.
It reveals that the anisotropic effects dominate at large angular scales (small \(\ell\))
while decay rapidly for small angular scales (large \(\ell\)).
On the other hand, the above result could also be revealed via the primordial power spectrum with anisotropy.
The anisotropic contribution in Eq.~(\ref{primordial power spectra of R final form}) is inversely proportional to the wavenumbers \(k\)'s.
Thus, it affects dominantly the modes at large scales (small \(k\)).
For small scales (large \(k\)), it decays rapidly.
These predictions imply that the statistical anisotropy discussed above affects only the low-\(\ell\) multipoles of the CMB anisotropy.

\section{V. Conclusions and remarks}
\label{conclusions}
In this paper, we resolved the quadrupole--octopole alignment of CMB anisotropy via the primordial power spectrum with anisotropy.
First, the inflationary universe was assumed to be of the Randers--Finsler type.
In the Randers spacetime, there is a privileged axis which characterizes the anisotropy of the spacetime.
We obtained the initial power spectrum with a dipolar modulation in this spacetime.
The dipolar modulation contributes only to the large--scale modes at the universe.
The ``power tensor'' technique was used to analyze the induced anisotropic effects.
In this model, however, the anisotropic contribution vanishes for the power tensor of a \(\ell\)--multipole.
Thus, the correlation matrix of power tensors was defined in our analysis.
This is a statistically significant quantity directly.
It describes the direct correlation between any two multipoles.
Thus, its eigenvectors with the maximum eigenvalue would lead to the privileged directions of alignments of the two multipoles.

After a lengthy calculation, we obtained the correlation matrix \(S_{ij}(2,3)\) for \(\ell=2\) and \(\ell'=3\).
It was found to be diagonal in the chosen coordinates.
It acquires the maximum eigenvalue at the plane perpendicular to the privileged axis in the Randers spacetime.
This result reveals that the quadrupole and octopole orientations would lie in this plane with the uniform distribution.
However, we cannot determine which direction is preferred for the alignment.
In the isotropic universe, by contrast, the quadrupole and octopole orientations are distributed uniformly at the celestial sphere.
Thus, the possibility for the quadrupole--octopole alignment is much larger in the anisotropic model than that in the isotropic model.
In addition, we noted that the anisotropic effect contributes only to the low--\(\ell\) multipoles.
Fig.~\ref{sl} shows that the correlation between two neighboring multipoles decays rapidly for large \(\ell\)'s.
This is consistent with the present observations.

Originally, the initial power spectrum with the dipolar modulation was used to resolve
the dipole modulation of CMB temperature fluctuations at large--angular scales.
Interestingly, the direction for the CMB dipole modulation was found to be nearly perpendicular to the one of the quadrupole--octopole alignment \cite{Planck2013resultsXXIII}.
There might be certain relevance between these two kinds of deviations from isotropy in the CMB.
Nevertheless, we need more precise observations to test this proposition in future, since the existed datasets have large uncertainties.
In addition, one should note that Finsler geometry is intrinsically anisotropic.
Any cosmological process would acquire similar anisotropic properties in Finsler spacetime.
Here we studied the inflationary phase of the universe in Randers spacetime.
We found that there could be certain imprints of such an anisotropic inflation on the CMB temperature fluctuations.
It would be interesting to test these imprints in future CMB observations.

\vspace{0.3 cm}
\begin{acknowledgments}
We are grateful to Yunguo Jiang, Danning Li, Ming-Hua Li, Hai-Nan Lin, and Bin Qin for useful help and discussions.
This work has been funded by the National Natural Science Fund of China under Grant No. 11075166 and No. 11147176.
\end{acknowledgments}

\section{Appendixes}
\begin{appendix}
\subsection{Appendix A. Calculation of the correlator \(C_{\ell\ell' m m'}\)}
\label{A}
In this appendix, we would like to give an explicit calculation of
the two-point correlator \(C_{\ell\ell' m m'}\).
One could also find a similar derivation in Ref.~\cite{Inflationary perturbations in anisotropic backgrounds2007}.
First, we could obtain the expression for \(a_{\ell m}\) by inverting Eq.~(\ref{CMB perturbation into spherical harmonics 1})
\begin{equation}
\label{CMB perturbation into spherical harmonics}
a_{\ell m}=\int d\Omega_{\hat{\textbf{p}}}\frac{\delta T}{T}(\hat{\textbf{p}},\textbf{x},\eta)Y^{*}_{\ell m}(\hat{\textbf{p}})\ .
\end{equation}
By substituting Eq.~(\ref{CMB perturbation into spherical harmonics}) into Eq.~(\ref{power spectrum--two point correlator}),
one obtains
\begin{equation}
\label{Cllmm}
C_{\ell \ell' m m'}=\int d\Omega_{\hat{\textbf{p}}} d\Omega_{\hat{\textbf{p}}'}
\langle \frac{\delta T}{T}(\hat{\textbf{p}},\textbf{x},\eta) \frac{\delta T^{*}}{T}(\hat{\textbf{p}}',\textbf{x},\eta) \rangle
Y^{*}_{\ell m}(\hat{\textbf{p}}) Y_{\ell' m'}(\hat{\textbf{p}}')\ .
\end{equation}
The CMB temperature fluctuations are evolved from the initial curvature perturbations
according to Eq.~(\ref{CMB perturbation given by primordial condition}).
Using Eq.~(\ref{CMB perturbation given by primordial condition}) and Eq.~(\ref{primordial power spectrum}),
we could get
\begin{equation}
\label{Cllmmdelta}
C_{\ell \ell' m m'}=\int \frac{d^3\textbf{k}}{(2\pi)^{3}}\frac{2\pi^{2}}{k^{3}}\mathcal{P}_{\mathcal{R}}(\textbf{k})
\int d\Omega_{\hat{\textbf{p}}} d\Omega_{\hat{\textbf{p}}'}
\Delta(k,\hat{\textbf{k}}\cdot\hat{\textbf{p}},\eta)\Delta^{*}(k,\hat{\textbf{k}}\cdot\hat{\textbf{p}}',\eta)
Y^{*}_{\ell m}(\hat{\textbf{p}}) Y_{\ell' m'}(\hat{\textbf{p}}')\ .
\end{equation}
The transfer function \(\Delta(k,\hat{\textbf{k}}\cdot\hat{\textbf{p}},\eta)\)
has been decomposed in terms of Legendre polynomials in Eq.~(\ref{transfer function}).
By substituting Eq.~(\ref{transfer function}) into Eq.~(\ref{Cllmmdelta}), one obtains
\begin{eqnarray}
\label{Cllmmdeltaell1}
C_{\ell \ell' m m'}&=&\int {d^3\textbf{k}}\frac{2\pi^{2}}{k^{3}}\mathcal{P}_{\mathcal{R}}(\textbf{k})
\sum_{\ell'' \ell'''}(-i)^{\ell''-\ell'''}\Delta_{\ell''}(k,\eta)\Delta^{*}_{\ell'''}(k,\eta)\nonumber\\
& &\times \int d\Omega_{\hat{\textbf{p}}} d\Omega_{\hat{\textbf{p}}'}
\left(\frac{2\ell''+1}{4\pi}\right)
P_{\ell''}(\hat{\textbf{k}}\cdot\hat{\textbf{p}})\left(\frac{2\ell'''+1}{4\pi}\right)
P^{*}_{\ell'''}(\hat{\textbf{k}}\cdot\hat{\textbf{p}}')
Y^{*}_{\ell m}(\hat{\textbf{p}}) Y_{\ell' m'}(\hat{\textbf{p}}')\ ,\\
\label{Cllmmdeltaell2}
&=&\int \frac{d^3\textbf{k}}{(2\pi)^{3}}\frac{2\pi^{2}}{k^{3}}\mathcal{P}_{\mathcal{R}}(\textbf{k})
\sum_{\ell'' m''}\sum_{\ell''' m'''}(-i)^{\ell''-\ell'''}\Delta_{\ell''}(k,\eta)\Delta^{*}_{\ell'''}(k,\eta)
Y^{*}_{\ell'' m''}(\hat{\textbf{k}})Y_{\ell''' m'''}(\hat{\textbf{k}})\nonumber\\
& &\times \int d\Omega_{\hat{\textbf{p}}} Y^{*}_{\ell m}(\hat{\textbf{p}})Y_{\ell'' m''}(\hat{\textbf{p}})
\int d\Omega_{\hat{\textbf{p}}'}Y^{*}_{\ell''' m'''}(\hat{\textbf{p}}')Y_{\ell' m'}(\hat{\textbf{p}}')\ ,\\
\label{Cllmmdeltaell3}
&=&\int \frac{d^3\textbf{k}}{(2\pi)^{3}}\frac{2\pi^{2}}{k^{3}}\mathcal{P}_{\mathcal{R}}(\textbf{k})
\left[(-i)^{\ell-\ell'}\Delta_{\ell}(k,\eta)\Delta^{*}_{\ell'}(k,\eta)
Y^{*}_{\ell m}(\hat{\textbf{k}})Y_{\ell' m'}(\hat{\textbf{k}})\right]\ .
\end{eqnarray}
Here Eq.~(\ref{Cllmmdeltaell3}) is just Eq.~(\ref{two point correlator}) in the main text.
From (\ref{Cllmmdeltaell1}) to (\ref{Cllmmdeltaell2}), we have used the fact that
Lengendre polynomial could be expanded in terms of the spherical harmonics, i.e.,
\begin{equation}
\label{Lengendre to spherical harmonic}
P_{\ell}(\hat{\textbf{k}}\cdot\hat{\textbf{p}})=\frac{4\pi}{2\ell+1}\sum_{m=-\ell}^{\ell}Y^{*}_{\ell m}(\hat{\textbf{k}})Y_{\ell m}(\hat{\textbf{p}})\ .
\end{equation}
From (\ref{Cllmmdeltaell2}) to (\ref{Cllmmdeltaell3}), we used the orthogonal relation of spherical harmonics, namely,
\begin{equation}
\label{orthogonality relation}
\int d\Omega_{\hat{\textbf{p}}} Y^{*}_{\ell m}(\hat{\textbf{p}})Y_{\ell' m'}(\hat{\textbf{p}})=\delta_{\ell \ell'}\delta_{m m'}\ .
\end{equation}
In the limit of statistical isotropy, Eq.~(\ref{Cllmmdeltaell3}) will reduce to the diagonal form,
\begin{equation}
\label{Cl}
C_{\ell\ell' m m'}^{iso}=C_{\ell}\delta_{\ell \ell'}\delta_{m m'}\ ,
\end{equation}
where the variance \(C_{\ell}\) is given by
\begin{equation}
\label{isotropic Cl}
C_{\ell}=\int \frac{dk}{k}\mathcal{P}_{\mathcal{R}}(k)|\Delta_{\ell}(k,\eta)|^2\ .
\end{equation}
This is just Eq.~(\ref{iso}) in the main text.

In the following, we discuss a special case where \(\langle A_{ij}(\ell) \rangle\) has no preferred directions
while the initial perturbations are anisotropic.
In this case, one should analyze the correlation matrix of power tensors to determine the possible alignment of two multipoles
as was discussed in Eq.~(\ref{power tensor correlation trace}) and Eq.~(\ref{power tensor correlation matrix}).
As an example, we discuss the simplest case that the initial power spectrum has a dipolar modulation, namely,
\cite{Direction dependence of the power spectrum1302,Inflation and primordial power spectra at anisotropic spacetime}
\begin{equation}
\label{g(k)}
\mathcal{P}_{\mathcal{R}}(\textbf{k})=\mathcal{P}^{iso}_{\mathcal{R}}(k)\left(1+g(k)(\hat{\textbf{k}}\cdot\hat{\textbf{n}})\right)\ ,
\end{equation}
where \(\mathcal{P}^{iso}_{\mathcal{R}}(k)\) denotes the isotropic part and \(g(k)\) is a function of \(k\).
Here \(\hat{\textbf{n}}\) denotes the privileged direction derived from certain anisotropic spacetime.
Let the third spatial axis \(x^3\) to align with \(\hat{\textbf{n}}\).
We could use the polar coordinate system in the momentum space.
The spherical harmonics are given as
\begin{equation}
\label{spherical harmonic}
Y_{\ell m}(\theta,\phi)=e^{im\phi}\sqrt{\frac{(2\ell+1)(\ell-m)!}{4\pi(\ell+m)!}}P^{m}_{\ell}(\cos\theta)\ ,
\end{equation}
where \(\theta\) is the angle between \(\hat{\textbf{k}}\) and \(z\)-axis.
The expression of \(\langle A_{ij}(\ell) \rangle\) could be written as
\begin{equation}
\label{Aijl expection}
\langle A_{ij}(\ell) \rangle=N(\ell)\sum_{m,m',m''}\langle \ell m|J_{i}|\ell m'\rangle\langle \ell m''|J_{j}|\ell m\rangle C_{\ell \ell m' m''}\ .
\end{equation}
To obtain \(C_{\ell \ell m' m''}\), one need only calculate the anisotropic effect as
\begin{eqnarray}
\label{anisotropic effect only}
C^{aniso}_{\ell \ell m' m''}&\equiv&C_{\ell \ell m' m''}-C_{\ell \ell m' m''}^{iso}\ \nonumber\\
&=&\int\frac{d k}{k}\mathcal{P}^{iso}_{\mathcal{R}}(k)g(k)|\Delta_{\ell}(k,\eta)|^{2} I\ ,
\end{eqnarray}
where \(I\) denotes the integral as
\begin{equation}
I=\int {d\Omega_{\hat{\textbf{k}}}}(\hat{\textbf{n}}\cdot\hat{\textbf{k}})Y^{*}_{\ell m'}(\hat{\textbf{k}})Y_{\ell m''}(\hat{\textbf{k}})\ .
\end{equation}
We could show that \(I\) always vanishes.
In the polar coordinates, \(I\) could be rewritten as
\begin{equation}
I=\int^{2\pi}_{0} d\phi \int^{\pi}_{0} \sin\theta d\theta \cos\theta Y^{*}_{\ell m'}(\theta,\phi)Y_{\ell m''}(\theta,\phi)\ .
\end{equation}
By substituting (\ref{spherical harmonic}) into the above equation, we obtain
\begin{equation}
I\propto {\delta_{m'm''}}\int ^{-1}_{1}d\zeta \zeta |P_{\ell}^{m'}(\zeta)|^2\ ,
\end{equation}
where \(\zeta=\cos\theta\).
It is obvious that the above integral always vanishes.
In this way, \(C_{\ell \ell m' m''}\) reduces to the isotropic case \(C_{\ell}\delta_{m' m''}\).
Thus, the dipolar modulation of the initial perturbations does not affect the eigenvalues of \(\langle A_{ij}(\ell) \rangle\).
Nevertheless, it could influence the correlation matrix of power tensors.
This issue has been discussed in Section II and Section IV.

\subsection{Appendix B. Expressions of \(J\) and \(J_{ij}\)}
\label{B}
In this Appendix, we give the explicit expressions of \(J\) and \(J_{ij}\).
The representation of \(\verb"so(3)"\) algebra is assumed in the following computation.
Especially, we use
\begin{eqnarray}
J_{\pm}&=&J_{1}\pm i J_{2}\ ,\\
J_{\pm}|\ell m \rangle&=&d_{\pm}(\ell,m)|\ell m \rangle\ ,\\
J_{3}|\ell m \rangle&=&m|\ell m \rangle\ ,
\end{eqnarray}
where
\begin{equation}
d_{\pm}(\ell,m)=\sqrt{(\ell\mp m)(\ell\pm m+1)}\ .
\end{equation}
According to the definition (\ref{J}), \(J\) could be obtained as \cite{Direction dependence of the power spectrum1302}
\begin{eqnarray}
\label{J expression}
J=\frac{1}{2}[d_{-}(\ell,m_{1})d_{+}(\ell',m'_{2})\delta_{m_{2},m_{1}-1}\delta_{m'_{1},m'_{2}+1}
&+&
d_{+}(\ell,m_{1})d_{-}(\ell',m'_{2})\delta_{m_{2},m_{1}+1}\delta_{m'_{1},m'_{2}-1}\nonumber\\
&+&2m_{1}m'_{1}\delta_{m_{1},m_{2}}\delta_{m'_{1},m'_{2}}]\ .
\end{eqnarray}
Similarly, the components of \(J_{ij}\) defined in (\ref{Jij}) are listed as follows:
\begin{eqnarray}
\label{J11}
J_{11}&=&\frac{1}{4}[d_{-}(\ell,m_{3})d_{+}(\ell',m'_{1})\delta_{m_{1},m_{3}-1}\delta_{m'_{3},m'_{1}+1}
+
d_{+}(\ell,m_{3})d_{-}(\ell',m'_{1})\delta_{m_{1},m_{3}+1}\delta_{m'_{3},m'_{1}-1}\nonumber\\
&&+
d_{-}(\ell,m_{3})d_{-}(\ell',m'_{1})\delta_{m_{1},m_{3}-1}\delta_{m'_{3},m'_{1}-1}
+
d_{+}(\ell,m_{3})d_{+}(\ell',m'_{1})\delta_{m_{1},m_{3}+1}\delta_{m'_{3},m'_{1}+1}]\ ,\\
\label{J22}
J_{22}&=&\frac{1}{4}[d_{-}(\ell,m_{3})d_{+}(\ell',m'_{1})\delta_{m_{1},m_{3}-1}\delta_{m'_{3},m'_{1}+1}
+
d_{+}(\ell,m_{3})d_{-}(\ell',m'_{1})\delta_{m_{1},m_{3}+1}\delta_{m'_{3},m'_{1}-1}\nonumber\\
&&-
d_{-}(\ell,m_{3})d_{-}(\ell',m'_{1})\delta_{m_{1},m_{3}-1}\delta_{m'_{3},m'_{1}-1}
-
d_{+}(\ell,m_{3})d_{+}(\ell',m'_{1})\delta_{m_{1},m_{3}+1}\delta_{m'_{3},m'_{1}+1}]\ ,\\
\label{J33}
J_{33}&=&m_{3}m'_{1}\delta_{m_{1},m_{3}}\delta_{m'_{1},m'_{3}}\ ,\\
\label{J12}
J_{12}&=&\frac{1}{4i}[d_{-}(\ell,m_{3})d_{+}(\ell',m'_{1})\delta_{m_{1},m_{3}-1}\delta_{m'_{3},m'_{1}+1}
-
d_{+}(\ell,m_{3})d_{-}(\ell',m'_{1})\delta_{m_{1},m_{3}+1}\delta_{m'_{3},m'_{1}-1}\nonumber\\
&&-
d_{-}(\ell,m_{3})d_{-}(\ell',m'_{1})\delta_{m_{1},m_{3}-1}\delta_{m'_{3},m'_{1}-1}
+
d_{+}(\ell,m_{3})d_{+}(\ell',m'_{1})\delta_{m_{1},m_{3}+1}\delta_{m'_{3},m'_{1}+1}]\ ,\\
\label{J21}
J_{21}&=&\frac{1}{4i}[d_{-}(\ell,m_{3})d_{+}(\ell',m'_{1})\delta_{m_{1},m_{3}-1}\delta_{m'_{3},m'_{1}+1}
-
d_{+}(\ell,m_{3})d_{-}(\ell',m'_{1})\delta_{m_{1},m_{3}+1}\delta_{m'_{3},m'_{1}-1}\nonumber\\
&&+
d_{-}(\ell,m_{3})d_{-}(\ell',m'_{1})\delta_{m_{1},m_{3}-1}\delta_{m'_{3},m'_{1}-1}
-
d_{+}(\ell,m_{3})d_{+}(\ell',m'_{1})\delta_{m_{1},m_{3}+1}\delta_{m'_{3},m'_{1}+1}]\ ,\\
\label{J13}
J_{13}&=&\frac{1}{2}m'_{1}\delta_{m'_{1},m'_{3}}[d_{-}(\ell,m_{3})\delta_{m_{1},m_{3}-1}+d_{+}(\ell,m_{3})\delta_{m_{1},m_{3}+1}]\ ,\\
\label{J31}
J_{31}&=&\frac{1}{2}m_{3}\delta_{m_{1},m_{3}}[d_{-}(\ell',m'_{1})\delta_{m'_{3},m'_{1}-1}+d_{+}(\ell',m'_{1})\delta_{m'_{3},m'_{1}+1}]\ ,\\
\label{J23}
J_{23}&=&\frac{1}{2i}m'_{1}\delta_{m'_{1},m'_{3}}[d_{+}(\ell,m_{3})\delta_{m_{1},m_{3}+1}-d_{-}(\ell,m_{3})\delta_{m_{1},m_{3}-1}]\ ,\\
\label{J32}
J_{32}&=&\frac{1}{2i}m_{3}\delta_{m_{1},m_{3}}[d_{+}(\ell',m'_{1})\delta_{m'_{3},m'_{1}+1}-d_{-}(\ell',m'_{1})\delta_{m'_{3},m'_{1}-1}]\ .
\end{eqnarray}
By substituting the above expressions of \(J\) and \(J_{ij}\) into Eq.~(\ref{correlation matrix 2}),
we could obtain the correlation matrix between any two neighboring multipoles in the Randers spacetime.

\end{appendix}


\begin{thebibliography}{999}

\bibitem{WMAP9001}C.L. Bennett {\it et al.},
    arXiv:1212.5225.

\bibitem{Planck2013resultsXXIII}Planck Collaboration, arXiv:1303.5083.

\bibitem{WMAP1001}C.L. Bennett {\it et al.},
    Astrophys. J. Suppl. {\bf 148}, 1 (2003).

\bibitem{quadrupole-octopole alignment}M. Tegmark, A. de Oliveira-Costa, A. Hamilton, Phys. Rev. D {\bf 68}, 123523 (2003).

\bibitem{quadrupole-octopole alignment by Schwarz}D.J. Schwarz, G.D. Starkman, D. Huterer, C.J. Copi, Phys. Rev. Lett. {\bf 93}, 221301 (2004).

\bibitem{quadrupole-octopole alignment by Copi1}C.J. Copi, D. Huterer, G.D. Starkman, Phys. Rev. D {\bf 70}, 043515 (2004).

\bibitem{quadrupole-octopole alignment by Copi}C.J. Copi, D. Huterer, D.J. Schwarz, G.D. Starkman, Mon. Not. Roy. Astron. Soc. {\bf 367}, 79 (2006).

\bibitem{hemispherical asymmetry01}H.K. Eriksen, F.K. Hansen, A.J. Banday, K.M. Gorski, P.B. Lilje, Astrophys. J. {\bf 605}, 14 (2004).

\bibitem{hemispherical asymmetry02}F.K.Hansen, A.J. Banday, K.M. Gorski, MNRAS {\bf 354}, 641 (2004).

\bibitem{hemispherical asymmetry03}C.-G. Park, MNRAS {\bf 349}, 313 (2004).

\bibitem{parity asymmetry01}K. Land, J. Magueijo, Phys. Rev. D, {\bf 72}, 101302 (2005).

\bibitem{parity asymmetry02}J. Kim, P. Naselsky, Astrophys. J. {\bf 714}, L265 (2010).

\bibitem{parity asymmetry03}P. Naselsky, W. Zhao, J. Kim, S. Chen, Astrophys. J. {\bf 749}, 31 (2012).

\bibitem{parity asymmetry04}J. Kim, P. Naselsky, Phys. Rev. D {\bf 82}, 063002 (2010).

\bibitem{parity asymmetry05}A. Gruppuso, et al., MNRAS {\bf 411}, 1445 (2011).

\bibitem{Dipole01}M.J. Longo, Phys. Lett. B {\bf 699}, 224 (2011).

\bibitem{Dipole02}L. Shamir, Phys. Lett. B {\bf 715}, 25 (2012).

\bibitem{parity asymmetry06}P.K. Aluri, P. Jain, MNRAS {\bf 419}, 3378 (2012).

\bibitem{Mirror parity01}V. Gurzadyan, A. Starobinsky, A. Kashin, H. Khachatryan, G. Yegorian, Mod. Phys. Lett. A {\bf 22}, 2955 (2007).

\bibitem{Mirror parity02}A. Ben-David, E.D. Kovetz, N. Itzhaki, Astrophys. J. {\bf 748}, 39 (2012).

\bibitem{Mirror parity03}F. Finelli, A. Gruppuso, F. Paci, A. Starobinsky, JCAP {\bf 07}, 049 (2012).

\bibitem{Directional dependence of CMB parity asymmetry}W. Zhao, arXiv:1306.0955.

\bibitem{Spontaneous isotropy breaking}C. Gordon, W. Hu, D. Huterer, T. Crawford, Phys. Rev. D {\bf 72}, 103002 (2005).

\bibitem{dipolar power modulation01}H.K. Eriksen, A.J. Banday, K.M. Gorski, F.K. Hansen, P.B. Lilje, ApJ {\bf 660}, L81 (2007).

\bibitem{dipolar power modulation02}J. Hoftuft {\it et al.}, Astrophys. J. {\bf 699}, 985 (2009).

\bibitem{dipolar power modulation03}D. Hanson, A. Lewis, Phys. Rev. D {\bf 80}, 063004 (2009).

\bibitem{The Virgo Alignment Puzzle in Propagation of Radiation on Cosmological Scales}J.P. Ralston, P. Jain, Int. J. Mod. Phys. D {\bf 13}, 1857 (2004).

\bibitem{Testing Isotropy of Cosmic Microwave Background Radiation}P.K. Samal, R. Saha, P. Jain, J.P. Ralston, Mon. Not. Roy. Astron. Soc. {\bf 385}, 1718 (2008).

\bibitem{Direction dependence of the power spectrum1302}P.K. Rath, T. Mudholkar, P. Jain, P.K. Aluri, S. Panda, JCAP {\bf 04}, 007 (2013).

\bibitem{Inflation and primordial power spectra at anisotropic spacetime}Z. Chang, S. Wang, Eur. Phys. J. C {\bf 73}, 2516 (2013).

\bibitem{Randers space}G. Randers,
    Phys. Rev. {\bf 59}, 195 (1941).

\bibitem{VSR}A.G. Cohen, S.L. Glashow,
    Phys. Rev. Lett. {\bf 97}, 021601 (2006).

\bibitem{Finsler isometry LiCM}X. Li, Z. Chang,
    Differ. Geom. Appl. {\bf 30}, 737 (2012).

\bibitem{The Finsler Type of Space-time Realization of Deformed Very Special Relativity}L. Zhang, X. Xue,
    arXiv:1205.1134.

\bibitem{Finsler}P. Finsler, {\it \(\ddot{U}\)ber Kurven and Fl\(\ddot{a}\)chen in allgemeinen R\(\ddot{a}\)umen}, (Dissertation, G\(\ddot{o}\)ttingan, 1918), Birkh\(\ddot{a}\)user Verlag, Basel, 1951.

\bibitem{Book by Bao}D. Bao, S.S. Chern, Z. Shen, {\it An Introduction to Riemann--Finsler Geometry},
        Graduate Texts in Mathematics {\bf 200}, Springer, New York, 2000.

\bibitem{Finsler isometry by Wang}H.C. Wang,
    J. London Math. Soc. {\bf s1-22} (1), 5 (1947).

\bibitem{Finsler isometry by Rutz}S.F. Rutz,
    Contemp. Math. {\bf 196}, 289 (1996).

\bibitem{A Hemispherical Power Asymmetry from Inflation}A.L. Erickcek, M. Kamionkowski, S.M. Carroll, Phys. Rev. D {\bf 78}, 123520 (2008).

\bibitem{The CMB asymmetry from inflation}D.H. Lyth, arXiv:1304.1270.

\bibitem{Small non-Gaussianity and dipole asymmetry in the CMB}L. Wang, A. Mazumdar, Phys. Rev. D {\bf 88}, 023512 (2013).

\bibitem{CMB dipole asymmetry from a fast roll phase}A. Mazumdar, L. Wang, arXiv:1306.5736.

\bibitem{Large Scale Anisotropic Bias from Primordial non-Gaussianity}S. Baghram, M.H. Namjoo, H. Firouzjahi, arXiv:1303.4368.

\bibitem{Hemispherical Asymmetry and Local non-Gaussianity a Consistency Condition}M.H.~Namjoo, S.~Baghram, H.~Firouzjahi, arXiv:1305.0813.


\bibitem{Asymmetric Sky from the Long Mode Modulations}A.A.~Abolhasani, S.~Baghram, H.~Firouzjahi and M.H.~Namjoo, arXiv:1306.6932.


\bibitem{Anisotropic Non-Gaussianity from a Two-Form Field}J. Ohashi, J. Soda, S. Tsujikawa, Phys. Rev. D {\bf 87}, 083520 (2013).

\bibitem{CMB Power Asymmetry from Primordial Sound Speed Parameter} Y.-F. Cai, W. Zhao, Y. Zhang, arXiv:1307.4090.

\bibitem{Book by Rund}H. Rund, {\it The Differential Geometry of Finsler Spaces}, Springer, Berlin, 1959.

\bibitem{Book by Asanov}G.S. Asanov, {\it Finsler Geometry, Relativity and Gauge Theories}, Reidel, Dordrecht (1995).

\bibitem{FRW model with weak anisotropy by Stavrinos}P.C. Stavrinos, A.P. Kouretsis, M. Stathakopoulos, Gen. Relativ. Gravit. {\bf 40}, 1403 (2008).


\bibitem{Inflation by Starobinsky}A.A. Starobinsky, Phys. Lett. B {\bf 91}, 99 (1980).

\bibitem{Inflation by Guth}A.H. Guth, Phys. Rev. D {\bf 23}, 347 (1981).

\bibitem{Inflation by Linde}A.D. Linde, Phys. Lett. B {\bf 108}, 389 (1982).

\bibitem{Inflation by Steinhardt}A. Albrecht, P.J. Steinhardt, Phys. Rev. Lett. {\bf 48}, 1220 (1982).

\bibitem{Inflation by Linde0}A.D. Linde, Phys. Lett. B {\bf 129}, 177 (1983).

\bibitem{Cosmological Magnetic Fields05}M. Watanabe, S. Kanno, J. Soda, Mon. Not. Roy. Astron. Soc.  {\bf 412}, L83 (2011).

\bibitem{Fluctuation01}V.F. Mukhanov, G. Chibisov, JETP Lett. {\bf 33}, 532 (1981).

\bibitem{Fluctuation02}V.F. Mukhanov, G. Chibisov, Sov. Phys. JETP {\bf 56}, 258 (1982).

\bibitem{Fluctuation03}S. Hawking, Phys. Lett. B {\bf 115}, 295 (1982).

\bibitem{Fluctuation04}A.H. Guth, S. Pi, Phys. Rev. Lett. {\bf 49}, 1110 (1982).

\bibitem{Fluctuation05}A.A. Starobinsky, Phys. Lett. B {\bf 117}, 175 (1982).

\bibitem{Fluctuation06}J.M. Bardeen, P.J. Steinhardt, M.S. Turner, Phys. Rev. D {\bf 28}, 679 (1983).

\bibitem{Fluctuation07}V.F. Mukhanov, JETP Lett. {\bf 41}, 493 (1985).

\bibitem{Particle physics models of inflation and curvaton scenarios}A. Mazumdar, J. Rocher, Phys. Rept. {\bf 497}, 85 (2011).

\bibitem{Visible sector inflation and the right thermal history in light of Planck data}L. Wang, E. Pukartas, A. Mazumdar, JCAP {\bf 07}, 019 (2013).

\bibitem{Inflationary perturbations in anisotropic backgrounds2007}A.E. Gumrukcuoglu, C.R. Contaldi, M. Peloso, JCAP {\bf 11}, 005 (2007).



\end{thebibliography}
\end{document}